\documentclass{article}

\usepackage{caption}

\catcode `\@=11
\@addtoreset{equation}{section}

\begin{document}

\def\refname{REFERENCES}

\hsize 140mm
\textheight 190mm

\title{\bf THE INFINITESIMAL-OPERATOR ALGEBRAS OF CONTINUOUS GROUPS 
WITH ANTILINEAR OPERATIONS}

\author{\bf J. Koci\'nski and M. Wierzbicki}

\maketitle

\begin{small}
\noindent
{\bf Abstract}.  Continuous groups with antilinear 
operations of the form $G+a_0G$, where $G$ denotes a linear Lie group, and
$a_0$ is an antilinear operation which fulfills the condition $a^2_0=\pm 1$, were defined
and their matrix algebras were investigated in \cite{Kocinski4}.  
In this paper infinitesimal-operator algebras
are defined for any group of the form $G+a_0G$, and their properties are determined.
\end{small}

\section{\bf Introduction}
\noindent
Birman's successful idea \cite{Birman} of introducing space groups with antilinear operations
to the description
of lattice vibrations suggested a definition of continuous groups
with antilinear operations of the form $G+a_0G$, where $G$ denotes a Lie group, and $a_0$
is an antilinear operation, fulfilling the condition $a^2_0=\pm 1$. The matrix algebras of
these groups were defined, their properties were investigated and examples were given
in \cite{Kocinski4}.
That investigation is extended in this paper, by defining
in Sections 4 and 5 the algebra of infinitesimal operators connected with
the groups $G+a_0G$. 
These are the operators $J_{\sigma}$, connected with the 
subgroup $G$, and the operators $J^{\prime}_{\mu}$, connected with the coset $a_0G$.
In the choice of the appropriate definitions we will be guided by the 
requirement that the one-to-one corespondence between the commutators of the infinitesimal
operators and the respective commutators of the 
basis vectors of the respective matrix algebra should hold.
There appear three types of commutators: $[J_{\sigma},\,J_{\tau}],
\,[J_{\sigma},\,J^{\prime}_{\mu}]$, and $[J^{\prime}_{\mu},\,J^{\prime}_{\nu}]$.
It will be shown that the commutators $[J_{\sigma},\,J_{\tau}]$ and the commutators
$[J^{\prime}_{\mu},\,J^{\prime}_{\nu}]$ yield linear combinations of operators 
connected with the subgroup $G$, while the commutators $[J_{\sigma},\,J^{\prime}_{\mu}]$
yield linear combinations of operators connected with the coset $a_0G$. 

\section{\bf Transformations of the coordinate space of corepresentations}.

\noindent
This section is based on the knowledge of the coirrep matrices of $a-$ and $b-$type coirreps
and of their basis functions, in the forms which were derived in \cite{Kocinski4}. 
We will denote the original coordinates
in the representation space of a corepresentation
by $y_j,\,j=1,...,d,d+1,...,2d$.
After the Kovalev-Gorbanyuk transformations of the corepresentation matrices in Eqs. (11)
and (12) of \cite{Kocinski4}, the respectively transformed coordinates acquire the following forms:
\\
\noindent
For $a-$type coirreps, after the application of transformation $V_1$ 
in Eq. (42) of \cite{Kocinski4}, we obtain the coordinate spaces of the reduced corepresentation
matrix in the form,
\begin{equation}
x^{(1)}_i=\frac{1}{\sqrt{2}}\Big(y_i+{\rm e}^{i\xi}\sum\limits_{j=1}^d N_{ij}y_{d+j}\Big),\quad
x^{(2)}_i=\frac{i}{\sqrt{2}}\Big(-y_i+{\rm e}^{i\xi}\sum\limits_{j=1}^d N_{ij}y_{d+j}\Big)
\quad
i=1,...,d
\label{eq2:1}
\end{equation}

\noindent
with $\mu/
\lambda ={\rm exp}(i\xi)$, with a real $\xi$, where the coordinates
$x^{(1)}_i,
\,i=1,...,d$ and $x^{(2)}_i,\,i=1,...,d$, are connected with 
the two $d-$dimensional blocks, respectively, 
with $N$ in Eqs. (20) and (39) of \cite{Kocinski4}.
\\
\noindent
For $b-$type coirreps, after the  application of the transformation $V_2$ in Eq. (52) of 
\cite{Kocinski4}, we obtain the coordinates in the space of the transformed
$2d-$dimensional matrices in the form:

\begin{equation}
x_i=-iy_i,\qquad x_{d+i}=-i\sum\limits_{k=1}^d N_{ik}y_{d+k},\quad i=1,...,d 
\label{eq2:2}
\end{equation}

\noindent
with $N$ in Eqs. (20) and (48) of \cite{Kocinski4}.
In the following we will omit the label "prime" of the 
corepresentation matrices $D^{\prime}(g)$ and $D^{\prime}(a)$, which was introduced in
\cite{Kocinski4}, remembering that they were obtained from the original
corepresentation matrices with the help of the transformations in 
Eqs. (42) and (52) of \cite{Kocinski4}. The coordinates in Eqs. (\ref{eq2:1}) and (\ref{eq2:2})
have to undergo another transformation, namely the transformation $S_1$ in Eq. (19) of
\cite{Kocinski4}, which introduces the factor ${\rm exp}(i\alpha_0)$ in front of the
coset $a_0G$ matrices. After that transformation the coordinates acquire the factor
${\rm exp}(i\alpha_0/2)$.
\\
\indent
We now can write out the results of action of the coirrep matrices in Eqs. (43) (44)
in \cite{Kocinski4} for
$a-$type coirreps, and in Eqs. (28), (50) and (51) in \cite{Kocinski4} for $b-$type coirreps,
supplied with 
the factor ${\rm exp}(i\alpha_0)$, on the respective cooordinates in Eqs. (\ref{eq2:1})
and (\ref{eq2:2}), supplied with the factor ${\rm exp}(i\alpha_0/2)$.
\\
\indent
{\em Coirreps of $a-$type}. Let $x^0$ denote a one-column matrix with the elements
$(x^0_1,...,x^0_d)$, of the type $x^{(1)}_i$ or $x^{(2)}_i$ in Eq. (\ref{eq2:1}), and let
$x^{\prime}$ denote a one-column matrix with the elements $x^{\prime}_1,...,x^{\prime}_d$.
The action of the subgroup $G$ matrices on the coordinates in the representation space
is determined by:

\begin{equation}
\Delta(g)\,{\rm exp}(i\alpha_0/2)x^0=x
\label{eq2:3}
\end{equation}

\noindent
where $x$ denotes a one-column matrix with the elements $(x_1,...,x_d)$. The respective
action of a single block of the coset matrix ${\rm exp}(i\alpha_0)D(ga_0)$, is determined by:

\begin{equation}
{\rm exp}(i\alpha_0)(\mu/\lambda)\Delta(g)N\,{\rm exp}(i\alpha_0/2)x^0=x^{\prime}
\label{eq2:4}
\end{equation}

\noindent
where $x^{\prime}$ denotes a one-column matrix, with the elements 
$(x^{\prime}_1,...,x^{\prime}_d)$. 
An analogous expression is obtained for the action of a single block
of the coset matrix ${\rm exp}(i\alpha_0)D(a_0g)$,

\begin{equation}
{\rm exp}(i\alpha_0)(\mu/\lambda)N\Delta^{\ast}(g)\,{\rm exp}(i\alpha_0/2)x^0=x^{\prime\prime}
\label{eq2:5}
\end{equation}

\noindent
where $x^{\prime\prime}$ denotes a one-column matrix, with the elements $x^{\prime\prime}_1,
...,x^{\prime\prime}_d$.
\\
\indent
{\em Coirreps of $b-$type}. Let

\begin{equation}
{\rm exp}(i\alpha_0/2)\left(\begin{array}{r}
x^0
\\
\hline
x^0_d
\end{array}\right)
\label{eq2:6}
\end{equation}

\noindent
denote a one-column matrix with the elements 
${\rm exp}(i\alpha_0/2)(x^0_1,...,x^0_d,x^0_{d+1},...,x^0_{2d})$
in its successive rows, where the first $d$ elements belong to $x^0$ and the remaining
$d$ elements to $x^0_d$. The transformations of that point with the matrices of the subgroup
$G$ are determined by:

\begin{equation}
\left(\begin{array}{r|r}
\Delta(g) & 0
\\
\hline
0 & \Delta(g)
\end{array}\right){\rm exp}(i\alpha_0/2)\left(\begin{array}{r}
x^0
\\
\hline
x^0_d
\end{array}\right)=\left(\begin{array}{r}
x
\\
\hline
x_d
\end{array}\right)
\label{eq2:7}
\end{equation} 

\noindent
where on the right hand side we have the one-column matrix with the elements of the first
$d$ rows denoted by $x$, and of the successive $d$ rows denote by $x_d$.
The transformations of the point in Eq. (\ref{eq2:6}) with the matrices $D(ga_0)$ and
$D(a_0g)$ of the coset $a_0G$ in Eqs. (50) and (51) of \cite{Kocinski4}, respectively, 
with the factor ${\rm exp}(i\alpha_0)$, are determined by:

\begin{equation}
{\rm exp}(i\alpha_0)\left(\begin{array}{r|r}
0 & \Delta(g)N
\\
\hline
-\Delta(g)N & 0
\end{array}\right)\,{\rm exp}(i\alpha_0/2)\left(\begin{array}{r}
x^0
\\
\hline
x^0_d
\end{array}\right)=\left(\begin{array}{r}
x^{\prime}_d
\\
\hline
x^{\prime}
\end{array}\right)
\label{eq2:8}
\end{equation}

\noindent
and
\begin{equation}
{\rm exp}(i\alpha_0)\left(\begin{array}{r|r}
0 & N\Delta^{\ast}(g)
\\
\hline
-N\Delta^{\ast}(g) & 0
\end{array}\right)\,{\rm exp}(i\alpha_0/2)\left(\begin{array}{r}
x^0
\\
\hline
x^0_d
\end{array}\right)=\left(\begin{array}{r}
x^{\prime\prime}_d
\\
\hline
x^{\prime\prime}
\end{array}\right)
\label{eq2:9}
\end{equation}

\section
{\bf The action of coirrep matrices on points in the coordinate space}

\noindent
Before defining the infinitesimal operators, we firstly have
to consider two types of products of corepresentation matrices:
(1) the product of two corep matrices of which each belongs to the coset $a_0G$, which is
equal to a matrix belonging to the subgroup $G$, and (2) the product of a matrix belonging
to the subgroup $G$ with a matrix belonging to the coset $a_0G$, which is equal to a matrix
belonging to the coset $a_0G$. The rules of
action of the corepresentation matrices on points in the corepresentation spaces have to be
established. We begin with the first type of product of two matrices.

\noindent
\hrulefill

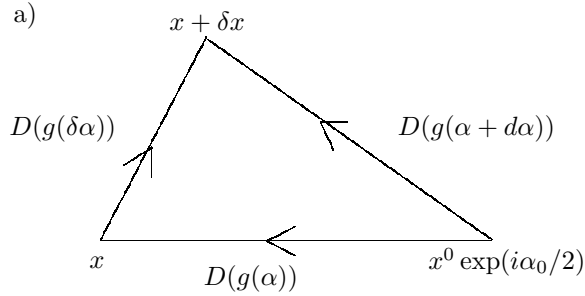
\begin{figure}[h]
\ifx\JPicScale\undefined\def\JPicScale{1}\fi
\unitlength \JPicScale mm
\begin{picture}(150,50)(0,0)
\linethickness{0.05mm}
\put(30,20){\line(1,0){52}}
\linethickness{0.05mm}
\multiput(44.17,46.74)(0.17,-0.12){223}{\line(1,0){0.17}}
\linethickness{0.05mm}
\multiput(30,20)(0.12,0.23){118}{\line(0,1){0.23}}
\put(29.49,17.13){\makebox(0,0)[cc]{$x$}}

\put(44.17,49.01){\makebox(0,0)[cc]{$x+\delta x$}}

\put(84,17.13){\makebox(0,0)[cc]{$x^0\exp(i\alpha_0/2)$}}

\linethickness{0.05mm}
\multiput(59.24,35.77)(0.12,-0.25){15}{\line(0,-1){0.25}}
\linethickness{0.05mm}
\put(59.24,35.77){\line(1,0){3.52}}
\linethickness{0.05mm}
\multiput(33,30)(0.19,0.12){20}{\line(1,0){0.19}}
\linethickness{0.05mm}
\put(36.73,28.37){\line(0,1){4.07}}
\linethickness{0.05mm}
\multiput(52,19.91)(0.23,0.12){17}{\line(1,0){0.23}}
\linethickness{0.05mm}
\multiput(52,19.91)(0.23,-0.12){17}{\line(1,0){0.23}}
\put(80,35){\makebox(0,0)[cc]{$D(g(\alpha+d\alpha))$}}

\put(65,35){\makebox(0,0)[cc]{}}

\put(25,35){\makebox(0,0)[cc]{$D(g(\delta\alpha))$}}

\put(50,15){\makebox(0,0)[cc]{$D(g(\alpha))$}}

\put(150,35){\makebox(0,0)[cc]{}}

\put(20,50){\makebox(0,0)[cc]{a)}}

\end{picture}
\vspace{-5em}
\caption*{Fig.1a The customary diagram for the Lie subgroup $G$ of the group $G+a_0G$.}
\end{figure}

\noindent
\hrulefill

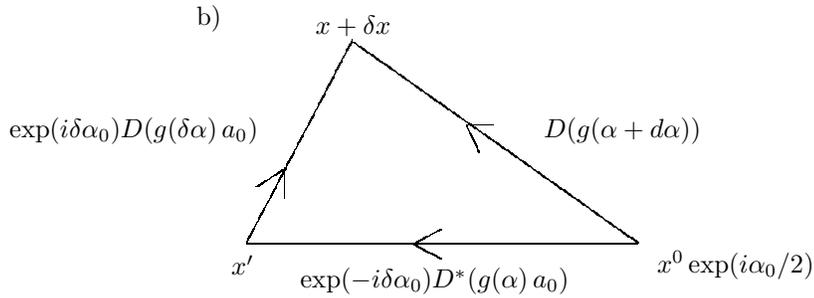
\begin{figure}[h]
\ifx\JPicScale\undefined\def\JPicScale{1}\fi
\unitlength \JPicScale mm
\begin{picture}(84,50)(0,0)
\linethickness{0.05mm}
\put(30,20){\line(1,0){52}}
\linethickness{0.05mm}
\multiput(44.17,46.74)(0.17,-0.12){223}{\line(1,0){0.17}}
\linethickness{0.05mm}
\multiput(30,20)(0.12,0.23){118}{\line(0,1){0.23}}
\put(29.49,17.13){\makebox(0,0)[cc]{$x'$}}

\put(44.17,49.01){\makebox(0,0)[cc]{$x+\delta x$}}

\put(95,17.13){\makebox(0,0)[cc]{$x^0\exp(i\alpha_0/2)$}}

\linethickness{0.05mm}
\multiput(59.24,35.77)(0.12,-0.25){15}{\line(0,-1){0.25}}
\linethickness{0.05mm}
\put(59.24,35.77){\line(1,0){3.52}}
\linethickness{0.05mm}
\multiput(31.27,27.56)(0.19,0.12){20}{\line(1,0){0.19}}
\linethickness{0.05mm}
\put(35,25.93){\line(0,1){4.07}}
\linethickness{0.05mm}
\multiput(52,19.91)(0.23,0.12){17}{\line(1,0){0.23}}
\linethickness{0.05mm}
\multiput(52,19.91)(0.23,-0.12){17}{\line(1,0){0.23}}
\put(80,35){\makebox(0,0)[cc]{$D(g(\alpha+d\alpha))$}}

\put(65,35){\makebox(0,0)[cc]{}}

\put(15,35){\makebox(0,0)[cc]{$\exp(i\delta\alpha_0)D(g(\delta\alpha)\,a_0)$}}

\put(55,15){\makebox(0,0)[cc]{$\exp(-i\delta\alpha_0)D^*(g(\alpha)\,a_0)$}}

\put(25,50){\makebox(0,0)[cc]{b)}}

\end{picture}
\vspace{-5em}
\caption*{Fig.1b The diagram for the coset $a_0G$ of the group $G+a_0G$.
This diagram is based on the property of the product of two corepresentation
matrices, each belonging to the coset $a_0G$, which always is equal to a matrix belonging
to the subgroup $G$.}
\end{figure}

\noindent
\hrulefill

\indent
The diagrams in Figs. 1a,b, refer to the corepresentation spaces
in Eqs. (\ref{eq2:1}) and (\ref{eq2:2}), 
of $a-$ or $b-$type coirreps, respectively.
\\
\indent
The diagram in Fig.~1a is the customary diagram for the Lie subgroup $G$ of the group
$G+a_0G$. In the diagram
in Fig.~1b we made use of the fact that the product of two corepresentation
matrices belonging to the coset $a_0G$ is equal to a corepresentation matrix belonging
to the subgroup $G$. 

\indent
The following argument is valid for $a-$ and $b-$type coirreps. For brevity, the symbols
$x^0$, $x$ and $x^{\prime}$ will represent the respective column matrices
for both $a-$type and $b-$type coirreps. 

In Fig. 1a, 
the point $x+\delta x$ in the corepresentation space is reached starting from the point 
$x^0{\rm exp}(i\alpha_0/2)$ by acting on it either with the transformation 
$D(g(\alpha+d\alpha))$, or
by acting on it successively with the transformations $D(g(\alpha))$ and $D(g(\delta\alpha))$,
all belonging to the subgroup $G$. This leads to the equality:

\begin{equation}
D(g(\alpha+d\alpha))x^0{\rm exp}(i\alpha_0/2)=
D(g(\delta\alpha))D(g(\alpha))x^0{\rm exp}(i\alpha_0/2)
\label{eq3:1}
\end{equation}

\noindent
As it is shown in Fig. 1b, the same point $x+\delta x$ can be obtained from the point
$x^0{\rm exp}i\alpha_0/2$ by acting on it either with the transformation $D(g(\alpha+d\alpha))$,
belonging to the subgroup $G$, or successively with
the transformations ${\rm exp}(-i\delta\alpha_0)D^{\ast}(g(\alpha)a_0)$ and 
${\rm exp}(i\delta\alpha_0)D(g(\delta\alpha)a_0)$, belonging to the coset $a_0G$. 
This leads to the equality:

\begin{equation}
D(g(\alpha+d\alpha))x^0{\rm exp}(i\alpha_0/2)=
{\rm exp}(i\delta\alpha_0)D(g(\delta\alpha)a_0){\rm exp}(-i\delta\alpha_0)
D^{\ast}(g(\alpha)a_0)x^0{\rm exp}(i\alpha_0/2)
\label{eq3:2}
\end{equation}

\noindent
These two equalities are valid for $a-$ and $b-$type coirreps.

\noindent
\hrulefill

\begin{figure}[h]
\ifx\JPicScale\undefined\def\JPicScale{1}\fi
\unitlength \JPicScale mm
\begin{picture}(150,50)(0,0)
\linethickness{0.05mm}
\put(30,20){\line(1,0){52}}
\linethickness{0.05mm}
\multiput(44.17,46.74)(0.17,-0.12){223}{\line(1,0){0.17}}
\linethickness{0.05mm}
\multiput(30,20)(0.12,0.23){118}{\line(0,1){0.23}}
\put(29.49,17.13){\makebox(0,0)[cc]{$\overline x$}}

\put(44.17,49.01){\makebox(0,0)[cc]{$\overline x+\delta \overline x$}}

\put(84,17.13){\makebox(0,0)[cc]{$x^0\exp(i\alpha_0/2)$}}

\linethickness{0.05mm}
\multiput(59.24,35.77)(0.12,-0.25){15}{\line(0,-1){0.25}}
\linethickness{0.05mm}
\put(59.24,35.77){\line(1,0){3.52}}
\linethickness{0.05mm}
\multiput(33,30)(0.19,0.12){20}{\line(1,0){0.19}}
\linethickness{0.05mm}
\put(36.73,28.37){\line(0,1){4.07}}
\linethickness{0.05mm}
\multiput(52,19.91)(0.23,0.12){17}{\line(1,0){0.23}}
\linethickness{0.05mm}
\multiput(52,19.91)(0.23,-0.12){17}{\line(1,0){0.23}}
\put(85,35){\makebox(0,0)[cc]{$\exp(i\delta\alpha_0)D(g(\alpha+d\alpha)a_0)$}}

\put(65,35){\makebox(0,0)[cc]{}}

\put(15,35){\makebox(0,0)[cc]{$\exp(i\delta\alpha_0)D(g(\delta\alpha)a_0)$}}

\put(50,15){\makebox(0,0)[cc]{$D^*(g(\alpha))$}}

\put(150,35){\makebox(0,0)[cc]{}}

\put(20,50){\makebox(0,0)[cc]{a)}}

\end{picture}
\vspace{-5em}
\caption*{Fig.2 The point $\overline x+\delta\,\overline x$
is obtained by acting on the point $x^0{\rm exp}(i\alpha_0/2)$,
with the transformation ${\rm exp}(i\delta\alpha_0)D(g(\alpha+d\alpha)a_0)$,
or with the successive transformations $D^{\ast}(g(\alpha))$, and 
${\rm exp(i\delta\alpha_0)}D(g(\delta\alpha)a_0)$.}
\end{figure}
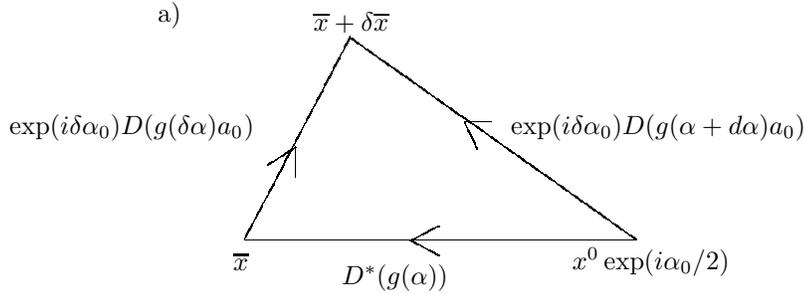

\noindent
\hrulefill

We next consider the second type of product, which is represented in Fig. 2.
In Fig. 2 we start from the point $x^0{\rm exp}(i\alpha_0/2)$ which is the same as in Figs. 1a,b.
Acting on this point with the matrix ${\rm exp}(i\delta\alpha_0)D(g(\alpha+d\alpha)a_0)$ we obtain the 
point $\overline x+\delta\,\overline x$. This point can alternatively be obtained after 
acting on the point
$x^0{\rm exp}(i\alpha_0/2)$ with the matrix $D^{\ast}(g(\alpha))$, obtaining the point 
$\overline x$, and next acting on the later point with the matrix 
${\rm exp}(i\delta\alpha_0)D(g(\delta\alpha)a_0)$.

We observe that the infinitesimal operators connected with the coset $a_0G$,
which can be defined at the point $\overline{x}$ in Fig. 2, are determined 
on the basis of the same transformation as the respective infinitesimal
operators which can be defined at the point $x'$ in Fig. 1b. This shows that
we can define the infinitesimal operators connected with the coset $a_0G$
at different points in the representation space. In the following calculations
we will define the infinitesimal operators connected with the coset $a_0G$ at the point $x'$.

We will establish the relations between the points $x^{\prime}$ and $x$. On the basis of Figs. 1a,b we define:

\begin{equation}
x=D(g(\alpha))x^0{\rm exp}(i\alpha_0/2),\qquad x^{\prime}={\rm exp}(-i\delta\alpha_0)D^{\ast}
(g(\alpha)a_0)x^0{\rm exp}(i\alpha_0/2)
\label{eq3:5}
\end{equation} 

\noindent
With these two definitions, the right hand sides of Eqs. (\ref{eq3:1}) and (\ref{eq3:2})
yield the relation between points $x$ and $x^{\prime}$:

\begin{equation}
\label{eq3:6}
D(g(\delta\alpha))x={\rm exp}(i\delta\alpha_0)D(g(\delta\alpha)a_0)x^{\prime}
\end{equation}

The explicit relations between the points $x^{\prime}$ and $x$ for $a-$ and
$b-$type coirreps will now be determined.
\\
\indent
{\em Coirreps of $a-$type}. These are given by a single block in (43) and (44) in \cite{Kocinski4} for the subgroup
$G$ and for the coset $a_0G$, respectively. The respective matrices are given in Eqs.
(\ref{eq2:3}), (\ref{eq2:4}) and (\ref{eq2:5}). The points $x$ and $x^{\prime}$
in Figs. 1a and 1b, respectively, are determined by the equalities

\begin{equation}
x=\Delta(g(\alpha))x^0{\rm exp}(i\alpha_0/2),\quad {\rm and}\quad 
x^{\prime}=(\mu/\lambda)^{\ast}\Delta^{\ast}(g(\alpha))N^{\ast}
x^0{\rm exp}(-i\alpha_0/2)
\label{eq3:9}
\end{equation}

\noindent
and the relation in Eq. (\ref{eq3:6}) between the points $x^{\prime}$ and $x$ takes the form:

\begin{equation}
\Delta(g(\delta\alpha))\,x=(\mu/\lambda)\Delta(g(\delta\alpha))N
{\rm exp}(i\delta\alpha_0)\,x^{\prime},\quad {\rm or}\quad x^{\prime}=N^{-1}{\rm exp}(-i\delta\alpha_0)\,x
\label{eq3:10}
\end{equation}

\noindent
since the factor $\mu/\lambda={\rm exp}(i\xi)$, with an arbitrary real $\xi$, can be absorbed
by the factor ${\rm exp}(i\delta\alpha_0)$.
\\
\indent
{\em Coirreps of $b-$type}. The coirrep matrices are  given in Eqs. (\ref{eq2:7}), (\ref{eq2:8}),
and (\ref{eq2:9}).
The points $x$ and $x^{\prime}$ in Figs. 1a and 1b, respectively, 
are determined by the equalities:

\begin{equation}
\left(\begin{array}{r}
x
\\
\hline
x_d
\end{array}\right)=
D(g(\alpha))\left(\begin{array}{r}
x^0
\\
\hline
x^0_d
\end{array}\right){\rm exp}(i\alpha_0/2)
\label{eq3:11}
\end{equation}

\noindent
and

\begin{equation}
\left(\begin{array}{r}
x^{\prime}_d
\\
\hline
x^{\prime}
\end{array}\right)=
{\rm exp}(-i\delta\alpha_0)D^{\ast}(g(\alpha)a_0)\left(\begin{array}{r}
x^0
\\
\hline
x^0_d
\end{array}\right){\rm exp}(i\alpha_0/2)
\label{eq3:12}
\end{equation}

\noindent
The relation between these two points, according to Eq. (\ref{eq3:6}) now takes the form:

\begin{equation}
D(g(\delta\alpha))\left(\begin{array}{r}
x
\\
\hline
x_d
\end{array}\right)={\rm exp}(i\delta\alpha_0)D(g(\delta\alpha)a_0)\left(\begin{array}{r}
x^{\prime}_d
\\
\hline
x^{\prime}\end{array}\right)
\label{eq3:13}
\end{equation}

\noindent
Substituting into this equality the expressions for the matrices $D(g(\delta\alpha))$ and
$D(g(\delta\alpha)a_0)$ from Eqs. (\ref{eq2:7}) and (\ref{eq2:8}), we obtain the equality:

\begin{equation}
\left(\begin{array}{r}
x^{\prime}
\\
\hline
x^{\prime}_d
\end{array}\right)={\rm exp}(-i\delta\alpha_0)\left(\begin{array}{r}
N^{-1}\,x
\\
\hline
-N^{-1}\,x_d
\end{array}\right)
\label{eq3:14}
\end{equation}

\section{\bf The infinitesimal operators of \textit{a}-type coirreps}

\noindent
We have to consider the infinitesimal operators $J_{\sigma}$ connected with the 
subgroup $G$, and $J^{\prime}_{\rho}$ connected with the
coset $a_0G$. The definitions of the infinitesimal operators connected with the subgroup $G$
and of their commutators are the same as those for the Lie subgroup $G$.
We distinguish by a "prime" the infinitesimal operators connected with the coset $a_0G$
from those connected with the subgroup $G$.
Two additional types of commutators: $[J^{\prime}_{\mu},J^{\prime}_{\nu}],
\,\mu,\nu=0,1,...,n$, and
$[J_{\sigma},J^{\prime}_{\mu}],\,\sigma=1,...,n,\,\mu=0,1,...,n$ have to be determined.
\\
\indent
The importance of the relation between the points $x$ and $x^{\prime}$
in Eq. (\ref{eq3:6}) is connected with the 
fact that the infinitesimal operators
for the subgroup $G$ will be defined at the point $x$, in accordance with Fig. 1a,
while the infinitesimal operators for the coset $a_0G$ will be defined at the point $x^{\prime}$
in accordance with Fig. 1b. We will require the equality of the commutator of two infinitesimal
operators connected with
the coset to a linear combination of operators connected with the subgroup, and the equality of
the commutator of an operator connected with the subgroup with an operator connected with 
the coset to a linear combination of operators connected with the coset.
The operators in the commutator have to be referred to the same point. 

\vspace{7mm}
\noindent
4.1 The subgroup $G$. 
\vspace{7mm}

\noindent
The definition  of the
infinitesimal operators $J_{\sigma},\,\sigma=1,...,n$, and the calculation of the
structural constants for the linear Lie subgroup $G$ of the group $G+a_0G$
carries over exactly from the Lie group theory.
The increments of the coordinates at the point $x$ in Fig. 1a, are given by

\begin{eqnarray}
dx_i=\Big(\frac{\partial}{\partial\alpha_{\sigma}}\Delta(g(\delta\alpha))_{ij}
\Big)_{\delta\alpha=0}x_j\delta\alpha_{\sigma}=(X_{\sigma})_{ij}x_j\delta\alpha_{\sigma}=
u_{i\sigma}(x)\delta\alpha_{\sigma}=u_{i\sigma}(x)
M^{-1}_{\sigma\lambda}d\alpha_{\lambda}
\nonumber\\
u_{i\sigma}(x)\equiv u_{i\sigma}(x_1,...,x_d),\quad i,j=1,...,d\quad\sigma=1,...,n
\label{eq4:1}
\end{eqnarray}

\noindent
with $\delta\alpha\equiv (\delta\alpha_1,...,\delta\alpha_n)$ and $x$ given in Eq. 
(\ref{eq2:3}),
where $(X_{\sigma})_{ij}$ is the element $(i,j)$ of the matrix $X_{\sigma}$ of the matrix basis
of the Lie subgroup $G$ algebra, and where 
the matrix $M_{\sigma\lambda}$ is determined
from the product of matrices $D(g(\alpha+d\alpha))=\Delta(g(\delta\alpha))\Delta(g(\alpha))$
according to Eq. (\ref{eq3:1}). In the customary way, we obtain the equations,

\begin{equation}
\frac{\partial x_i}{\partial\alpha_{\lambda}}=u_{i,\sigma}(x)M^{-1}_{\sigma\lambda},
\quad i=1,...,d;\quad \lambda,\sigma=1,...,n 
\label{eq4:2}
\end{equation}

\noindent
From the integrability condition of these equations we derive 
the structural constants $c^{\tau}_{\sigma\rho}$. Defining the infinitesimal operators:
\begin{equation}
J_{\sigma}=u_{i\sigma}(x)\frac{\partial}{\partial x_i}=(X_{\sigma})_{ij}x_j
\frac{\partial}{\partial x_i},\qquad i,k=1,...,d,\quad\sigma=1,...,n
\label{eq4:3}
\end{equation}

\noindent
where $(X_{\sigma})_{ij}$ is the $(ij)-$element of the matrix basis vector
$X_{\sigma}$, we determine the commutators

\begin{equation}
[J_{\sigma},J_{\rho}]=c^{\tau}_{\sigma\rho}J_{\tau}
\label{eq4:4}
\end{equation}

\noindent
as in the Lie groups, \cite{Cornwell,Hammermesh, Naimark}.

\vspace{7mm}
\noindent
4.2 The coset $a_0G$. 
\vspace{7mm}

\noindent
Since we have $a^2_0=\pm 1$, 
the matrix $\Delta (a^2_0)$ in (38) of \cite{Kocinski4} is equal to $\pm E$,
and then $\mu/\lambda={\rm exp (i\xi)}$, with a real $\xi$, which can be absorbed by
${\rm exp}(i\delta\alpha_0)$.
Performing an infinitesimal transformation of the point $x^{\prime}$, which was defined in
Fig.1b, with
the matrix ${\rm exp}(i\delta\alpha_0)\Delta(g(\delta\alpha))N$, we obtain:

\begin{eqnarray}
dx^{\prime}_i=\Big(\frac{\partial}{\partial\alpha_{\sigma}}{\rm exp}(i\delta\alpha_0)\Delta 
(g(\delta\alpha))_{ij}\Big)_{\delta\alpha=0}N_{jk}x^{\prime}_k
\delta\alpha_{\sigma}=(X^{\prime}_{\sigma})_{ik}x^{\prime}_k\delta\alpha_{\sigma}=
\nonumber\\
u^{\prime}_{i\sigma}(x^{\prime})\delta\alpha_{\sigma}
\quad i,j,k=1,...,d;\quad\sigma=0,1,...,n
\label{eq4:5}
\end{eqnarray}

\noindent
where the derivatives are calculated at the point $\delta\alpha_0=\delta\alpha_1=...=
\delta\alpha_n=0$, and $(X^{\prime}_{\sigma})_{ik}$ is the $(ik)-$element of the matrix
basis vector $X^{\prime}_{\sigma}$ connected with the coset $a_0G$, which was defined in
\cite{Kocinski4}.
\\
\indent
We next define the infinitesimal operators connected with the coset $a_0G$:
$J^{\prime}_{\sigma},\,\sigma=0,1,...,n,$ at the point $x^{\prime}$ (see Fig.1b),
in the form:

\begin{equation}
J^{\prime}_{\sigma}=u^{\prime}_{i\sigma}(x^{\prime})\frac{\partial}{\partial x^{\prime}_i}=(X^{\prime}_{\sigma})_{ik}x^{\prime}_k\frac{\partial}
{\partial x^{\prime}_i},
\qquad\sigma=0,1,...,n;\quad i,k=1,...,d
\label{eq4:6}
\end{equation}

\noindent
according to Eq.(\ref{eq4:5}).
\\
\indent
The commutators  $[J^{\prime}_{\sigma},\,J^{\prime}_{\rho}],\,\sigma\neq
\rho=0,1,...,n,$ have the form: 

\begin{eqnarray}
[J^{\prime}_{\sigma},\,J^{\prime}_{\rho}]=\Big[(X^{\prime}_{\sigma})_{ik}x^{\prime}_k
\frac{\partial}{\partial x^{\prime}_i},\,(X^{\prime}_{\rho})_{jl}x^{\prime}_l
\frac{\partial}{\partial x^{\prime}_j}\Big]=
\nonumber\\
\sum\limits_{i,j,k,l=1}^d(X^{\prime}_{\sigma})_{ik}(X^{\prime}_{\rho})_{jl}
\Big(\,x^{\prime}_k\frac{\partial}
{\partial x^{\prime}_i}x^{\prime}_l\frac{\partial}{\partial x^{\prime}_j}-
x^{\prime}_l\frac{\partial}{\partial x^{\prime}_j}x^{\prime}_k
\frac{\partial}{\partial x^{\prime}_i}\Big)
\label{eq4:7}
\end{eqnarray}

\noindent
After the transformation to the point $x$,
according to Eq. (\ref{eq3:10}), the commutator acquires the form of a linear combination of infinitesimal operators connected with the subgroup $G$.
\\
\indent 
For the remaining type of commutators we find:

\begin{eqnarray}
[J_{\sigma},\,J^{\prime}_{\rho}]=\Big[(X_{\sigma})_{ik}x_k\frac{\partial}{\partial x_i},
\,(X^{\prime}_{\rho})_{jl}x^{\prime}_l
\frac{\partial}{\partial x^{\prime}_j}\Big]=
\nonumber\\
\sum\limits_{i,j,k,l=1}^d(X_{\sigma})_{ik}(X^{\prime}_{\rho})_{jl}\Big(x_k
\frac{\partial}{\partial x_i}x^{\prime}_l\frac{\partial}
{\partial x^{\prime}_j}-x^{\prime}_l\,\frac{\partial}{\partial x^{\prime}_j}
x_k\frac{\partial}{\partial x_i}\Big)
\nonumber\\
\sigma=1,...,n;\quad \rho=0,1,...,n
\label{eq4:8}
\end{eqnarray}

\noindent
After the transformation of the terms depending
on the point $x$ to the point $x^{\prime}$, we obtain
a linear combination of infinitesimal operators connected with the coset $a_0G$.

\section{\bf The infinitesimal operators of {\em b}-type coirreps.}

\noindent
5.1 The subgroup G.

\vspace{5mm}
\noindent
As the corep matrices are constructed from square blocks of equal dimensions, it will be 
convenient to label the coordinates from 1 to $d$ and from $d+1$ to $2d$, where $d$ denotes
the dimension of the irrep $\Gamma$ of the subgroup $G$, with matrices $\Delta (g)$.
The transformation of the point $(x|x_d)$ in Eq. (\ref{eq2:7}) with the 
matrix $D(g(\delta\alpha))$ is determined by

\begin{equation}
\left(\begin{array}{r|r}
\Delta(g(\delta\alpha)) & 0
\\
\hline
0 & \Delta(g(\delta\alpha)) 
\end{array}\right)\left(\begin{array}{r}
x
\\
\hline
x_d
\end{array}\right)=\left(\begin{array}{r}
dx
\\
\hline
dx_d
\end{array}\right)
\label{eq5:1}
\end{equation}

\noindent
The structural constants and the infinitesimal operators connected with
the subgroup $G$, are obtained from the infinitesimal increments of the coordinates
at the point $(x_1,...,x_d,x_{d+1},...,x_{2d})$, in the form:

\begin{eqnarray}
dx_i=\Big(\frac{\partial}
{\partial\alpha_{\sigma}}\Delta (g(\delta\alpha))_{ij}
\Big)_{\delta\alpha=0}x_j\delta\alpha_{\sigma}=(\overline X_{\sigma})_{ij}
x_j\delta\alpha_{\sigma}=
u_{i\sigma}(x_w)\delta\alpha_{\sigma}=
u_{i\sigma}(x_w)M^{-1}_{\sigma\lambda}d\alpha_{\lambda}
\nonumber\\
u_{i\sigma}(x_w)\equiv u_{i\sigma}(x_1,...,x_d);\quad
i,j=1,...,d;\quad \sigma, \lambda=1,...,n
\label{eq5:2}
\end{eqnarray}

\noindent
and
\begin{eqnarray}
dx_{d+i}=\Big(\frac{\partial}{\partial\alpha_{\sigma}}
\Delta (g(\delta\alpha))_{ij}\Big)_{\delta\alpha=0}x_{d+j}\delta\alpha_{\sigma}=
(\overline X_{\sigma})_{ij}x_{d+j}\delta\alpha_{\sigma}=
u_{d+i,\sigma}(x_{d+w})\delta\alpha_{\sigma}=
\nonumber\\
u_{d+i,\sigma}(x_{d+w})M^{-1}_{\sigma\lambda}d\alpha_{\lambda};\qquad
u_{d+i,\sigma}(x_{d+w})\equiv u_{d+i,\sigma}(x_{d+1},...,x_{2d})
\nonumber\\
i,j=1,...,d;\quad \sigma, \lambda=1,...,n
\label{eq5:3}
\end{eqnarray}

\noindent
where the derivatives with respect to $\alpha_{\sigma}$ are calculated at $\delta\alpha=
(\delta\alpha_1,...,\delta\alpha_n)=0$, and where $(\overline X_{\sigma})_{ij}$ is the element
$(i,j)$ of the upper non-zero block of the matrix $X_{\sigma}$, belonging to the matrix
basis of the subgroup $G$ algebra. 
The matrix $M_{\lambda\sigma}$ is calculated
as in the case of $a-$type coirreps. We obtain the equations,

\begin{eqnarray}
\frac{\partial x_i}{\partial\alpha_{\lambda}}=u_{i\sigma}(x)M^{-1}_{\sigma\lambda},
\quad {\rm and}\quad \frac{\partial x_{d+i}}{\partial\alpha_{\lambda}}=u_{d+i,\sigma}(x_{w+d})
M^{-1}_{\sigma\lambda}
\nonumber\\
i=1,...,d;\quad \sigma,\lambda=1,...,n
\label{eq5:4}
\end{eqnarray}

\noindent
From the integrability conditions of these equations the respective
structural constants $c^{\tau}_{\sigma\rho}$ are derived. The infinitesimal
operators are define by:

\begin{eqnarray}
J_{\sigma}=u_{i\sigma}(x_1,...,x_d)\frac{\partial}{\partial x_i}+
u_{d+i,\sigma}(x_{d+1},...,x_{2d})\frac{\partial}{\partial x_{d+i}}=
\nonumber\\
(\overline X_{\sigma})_{ij}x_j\frac{\partial}{\partial x_i}+
(\overline X_{\sigma})_{ij}x_{d+j}\frac{\partial}{\partial x_{d+i}},\qquad
i,j=1,...,d;\quad\sigma=1,...,n
\label{eq5:5}
\end{eqnarray}

\noindent
For the commutator of two infinitesimal operators $J_{\sigma}$ and $J_{\rho}$,
connected with the subgroup $G$, we find in the way analogous to that for the $a-$type coirreps
the expression:

\begin{equation}
[J_{\sigma},J_{\rho}]=c^{\tau}_{\sigma\rho}J_{\tau};\qquad\sigma,\rho,\tau=1,...,n
\label{eq5:6}
\end{equation}

\noindent
with the structural constants $c^{\tau}_{\sigma\rho}$.

\vspace{7mm}
\noindent
5.2 The coset $a_0G$.
\vspace{7mm}

\noindent
The transformation of the point $(x^{\prime}_d|x^{\prime})$ in Eq. (\ref{eq3:12}) with the  matrix
${\rm exp}(i\delta\alpha_0)D(g(\delta\alpha)a_0)$ has the form:

\begin{eqnarray}
{\rm exp}(i\delta\alpha_0)\left(\begin{array}{r|r}
0 & \Delta(g(\delta\alpha))N
\\
\hline
-\Delta(g(\delta\alpha))N & 0
\end{array}\right)\left(\begin{array}{r}
x^{\prime}_d
\\
\hline
x^{\prime}
\end{array}\right)=
\nonumber\\
{\rm exp}(i\delta\alpha_0)\left(\begin{array}{r}
\Delta(g(\delta\alpha))Nx^{\prime}
\\
\hline
-\Delta(g(\delta\alpha))Nx^{\prime}_d
\end{array}\right)=
\left(\begin{array}{r}
dx^{\prime}
\\
\hline
dx^{\prime}_d
\end{array}\right)
\label{eq5:7}
\end{eqnarray}

\noindent
from which we obtain the expressions:

\begin{eqnarray}
dx^{\prime}_i=\Big(\frac{\partial}{\partial
\alpha_{\sigma}}{\rm exp}(i\delta\alpha_0)\Delta (g(\delta\alpha))_{ij}\Big)_{\delta\alpha=0}
N_{jk}x^{\prime}_k\delta\alpha_{\sigma}=({\overline X^{\,\prime}_{\sigma}})_{ik}x^{\prime}_k
\delta\alpha_{\sigma}=
\nonumber\\
u^{\prime}_{i\sigma}(x^{\prime}_1,...,x^{\prime}_d)\delta\alpha_{\sigma};\qquad
i,j,k=1,...,d,\quad\lambda,\sigma=0,1,...,n
\label{eq5:8}
\end{eqnarray}

\noindent
where 
$\overline X^{\,\prime}_{\sigma}$ denotes the upper non-zero block of the 
matrix $X^{\prime}_{\sigma}$ belonging to the matrix basis of the $b-$type coirrep algebra, and

\begin{eqnarray}
dx^{\prime}_{d+i}=-\Big(\frac{\partial}{\partial\alpha_{\sigma}}
{\rm exp}(i\delta\alpha_0)\Delta (g(\delta\alpha))_{ij}
\Big)_{\delta\alpha=0}N_{jk}x^{\prime}_{d+k}\delta\alpha_{\sigma}=
-(\overline X^{\,\prime}_{\sigma})_{ik}x^{\prime}_{d+k}\delta\alpha_{\sigma}=
\nonumber\\
u^{\prime}_{d+i,\sigma}(x^{\prime}_{d+1},...,x^{\prime}_{2d})\delta\alpha_{\sigma}
\qquad i,j,k=1,...,d,\quad\lambda,\sigma=0,1,...,n
\label{eq5:9}
\end{eqnarray}

\noindent
where in Eqs. (\ref{eq5:8}) and (\ref{eq5:9}) the derivatives are calculated at the point:
$\delta\alpha=(\delta\alpha_0,\delta\alpha_1,...,\delta\alpha_n)=0$. 
For brevity we will define:

\begin{eqnarray}
(x^{\prime}_1,...,x^{\prime}_d)\equiv x^{\prime}_w,\quad {\rm and}\quad (x^{\prime}_{d+1},
...,x^{\prime}_{2d})\equiv x^{\prime}_{d+w}
\nonumber\\
u^{\prime}_{i\sigma}(x^{\prime}_1,..,x^{\prime}_d)\equiv u^{\prime}_{i\sigma}(x^{\prime}_w),\quad {\rm and}\quad
u^{\prime}_{d+i,\sigma}(x^{\prime}_{d+1},...,x^{\prime}_{2d})\equiv
u^{\prime}_{d+i,\sigma}(x^{\prime}_{d+w})
\label{eq5:10}
\end{eqnarray}

\vspace{7mm}
\noindent
The infinitesimal operators $J^{\prime}_{\sigma},\,\sigma=0,1,...,n$, connected with
the coset $a_0G$ are defined at the point $(x^{\prime}|x^{\prime}_d)$ in Eq. (\ref{eq3:12})
in the form:

\begin{eqnarray}
J^{\prime}_{\sigma}=u^{\prime}_{i\sigma}(x^{\prime}_w)\frac{\partial}{\partial x^{\prime}_i}+
u^{\prime}_{d+i,\sigma}(x^{\prime}_{d+w})\frac{\partial}{\partial x^{\prime}_{d+i}}
=(\overline X^{\,\prime}_{\sigma})_{ik}\,x^{\prime}_k\frac{\partial}{\partial x^{\prime}_i}-(\overline X^{\,\prime}_{\sigma})_{ik}\,x^{\prime}_{d+k}\frac
{\partial}{\partial x^{\prime}_{d+i}}
\nonumber\\
i=1,...,d;\quad\sigma=0,1,...,n
\label{eq5:11}
\end{eqnarray}

\noindent
For the commutators $[J^{\prime}_{\sigma},\,J^{\prime}_{\rho}]$, with
$\sigma=1,...,n,\,\rho=0,1,...,n$
we find the following expression:

\begin{eqnarray}
[J^{\prime}_{\sigma},\,J^{\prime}_{\rho}]=
\Big[(\overline X^{\,\prime}_{\sigma})_{ik}
\Big(x^{\prime}_k
\frac{\partial}{\partial x^{\prime}_i}-x^{\prime}_{d+k}\frac{\partial}
{\partial x^{\prime}_{d+i}}\Big),\,(\overline X^{\,\prime}_{\rho})_{jl}
\Big(x^{\prime}_l\frac{\partial}{\partial x^{\prime}_j}-x^{\prime}_{d+l}
\frac{\partial}{\partial x^{\prime}_{d+j}}\Big)\Big]=
\nonumber\\
\sum\limits_{i,j,k,l=1}^d(\overline X^{\,\prime}_{\sigma})_{ij}(\overline X^{\,\prime}_{\rho})_{jl}\Big(x^{\prime}_k
\frac{\partial}{\partial x^{\prime}_i}x^{\prime}_l\frac{\partial}
{\partial x^{\prime}_j}-x^{\prime}_l\frac{\partial}{\partial x^{\prime}_j}
x^{\prime}_k\frac{\partial}{\partial x^{\prime}_i}\Big)+
\nonumber\\
\sum\limits_{i,j,k,l=1}^d(\overline X^{\,\prime}_{\sigma})_{ik}(\overline X^{\,\prime}_{\rho})_{jl}\Big(x^{\prime}_{d+k}\frac{\partial}
{\partial x^{\prime}_{d+i}}x^{\prime}_{d+l}\frac{\partial}
{\partial x^{\prime}_{d+j}}-x^{\prime}_{d+l}\frac{\partial}
{\partial x^{\prime}_{d+j}}x^{\prime}_{d+k}\frac{\partial}
{\partial x^{\prime}_{d+i}}\Big)
\label{eq5:12}
\end{eqnarray}

\noindent
After the transformation to the point $x$, the last expression turns into a linear
combination of the infinitesimal operators connected with the subgroup $G$. 
\\
\indent
The remaining type of commutators has the form:
$[J_{\sigma},\,J^{\prime}_{\rho}],\, \sigma=1,...,n,\,\rho=0,1,...,n$. Introducing for
$J_{\sigma}$ and $J^{\prime}_{\rho}$ the expressions in Eqs. (\ref{eq5:5}) and (\ref{eq5:11})
respectively, we obtain:

\begin{eqnarray}
[J_{\sigma},\,J^{\prime}_{\rho}]=
\Big[(\overline X_{\sigma})_{ik}\Big(x_k\frac{\partial}{\partial x_i}+
x_{d+k}\frac{\partial}{\partial x_{d+i}}\Big),\,(\overline X^{\,\prime})_{jl}
\Big(x^{\prime}_l\frac{\partial}{\partial x^{\prime}_j}-x^{\prime}_{d+l}
\frac{\partial}{\partial x^{\prime}_{d+j}}\Big)\Big]=
\nonumber\\
\sum\limits_{i,j,k,l=1}^d(\overline X_{\sigma})_{ik}(\overline X^{\,\prime}_{\rho})_{jl}\Big(x_k\frac{\partial}{\partial x_i}\,x^{\prime}_l
\frac{\partial}{\partial x^{\prime}_j}-x^{\prime}_l
\frac{\partial}{\partial x^{\prime}_j}x_k\frac{\partial}{\partial x_i}\Big)+
\nonumber\\
\sum\limits_{i,j,k,l=1}^d(\overline X_{\sigma})_{ik})(\overline X^{\prime}_{\rho})_{jl}
\Big(x_{d+k}\frac{\partial}
{\partial x_{d+i}}\,x^{\prime}_{d+l}\frac{\partial}{\partial x^{\prime}_{d+j}}
-x^{\prime}_{d+l}\frac{\partial}{\partial x^{\prime}_{d+j}}
x_{d+k}\frac{\partial}{\partial x_{d+i}}\Big)
\label{eq5:13}
\end{eqnarray}

\noindent
After the transformation to the point
$x^{\prime}$ in Fig. 1b, the last expression turns into 
a linear combination of operators defined in Eq. (\ref{eq5:11}).

\vspace{7mm}
\noindent
{\bf Observation 5.1}. When the operators $J^{\prime}_{\rho}$ are linearly independent on
the operators $J_{\sigma},\,\sigma=1,...,n,\, \rho=0,1,...,n$, which always holds
for $b$-type coirreps, the real algebra connected with the group
$G+a_0G$ is spanned by the the operators $J_{\sigma}$ and $J^{\prime}_{\rho}$, and it 
is $(2n+1)-$dimensional. When for $a-$type coirreps, the operators $J^{\prime}_{\rho},\,\rho=1,...,n$, 
linearly depend on the operators $J_{\sigma}$,
the respective real algebra is spanned by the operators
$J_{\sigma},\,\sigma=1,...,n$ together with the operator $J^{\prime}_0$, and it is
$(n+1)-$dimensional.

\section{\bf Conclusions}

\noindent
The matrix algebras of continuous groups with antilinear operations of the type
$G+a_0G$, where $G$ denotes a Lie group and the antilinear operation $a_0$ fulfills
the condition $a^2_0=\pm 1$, were determined in \cite{Kocinski4}. In this paper the infinitesimal operators of these groups were determined, when the coirreps of
the groups $G+a_0G$ are of $a-$type or of $b-$type. The infinitesimal operators connected with the subgroup $G$ differ from those connected with the coset $a_0G$.
By definition the commutators of the infinitesimal operators are to be compatible
with the commutators of the basis matrices of the matrix algebra which was determined in \cite{Kocinski4}.

\end{document}